# Infrared signatures of charge stripes in La$_{2-x}$Sr$_x$CuO$_4$.

A. Lucarelli[1], S. Lupi[1], M. Ortolani[1], P. Calvani[1], P. Maselli[1], M. Capizzi[1], P. Giura[1]†, H. Eisaki[2], N. Kikugawa[2], T. Fujita[3], M. Fujita[4], K. Yamada[4]

[1] *Istituto Nazionale di Fisica della Materia  and Dipartimento di Fisica, Università  di Roma La Sapienza, Piazzale Aldo Moro 2, I-00185 Roma, Italy*

[2] *Laboratory for Advanced  Materials, Stanford University, Stanford CA94305, U. S. A.*

[3] *ADSM, Hiroshima University, Higashi-Hiroshima 739-8526, Japan*

[4] *Institute for Chemical Research, Kyoto University, Gokasho, Uji 610-0011, Japan*

†Present address: Inelastic Scattering Group, ESRF, BP 220, Grenoble 38043, France

*Abstract: The in-plane optical conductivity of seven La$_{2-x}$Sr$_x$CuO$_4$ single crystals with $0 \leq x \leq 0.15$ has been studied between 30 and 295 K.  All doped samples exhibit strong peaks in the far-infrared, which closely resemble those observed in Cu-O "ladders" with one-dimensional charge-ordering. The behavior with doping and temperature of the peak energy, width, and intensity allows us to conclude that we are observing  the infrared absorption of charge stripes in La$_{2-x}$Sr$_x$CuO$_4$.*

According to theoretical predictions,[1-3] the charges injected by doping into the Cu-O planes of high-T$_c$ superconductors may gather into fluctuating, one-dimensional structures (stripes) below a temperature T* >> T$_c$. Their existence is confirmed indirectly by the magnetic order[4,5] attributed to the regions between the stripes, by lattice distortions detected in the Extended  X-ray Absorption Fine Structure,[6] or by the opening at T* of "pseudogaps" in the density of states of the carriers.[7] However, the excitation of charge arrays in the polar Cu-O lattice should also produce strong dipole fluctuations that could be directly detected by infrared spectroscopy.

The phase diagram of La$_{2-x}$Sr$_x$CuO$_4$ (LSCO) is recalled in the inset of Fig. 3. The insulator-to-superconductor transition occurs[8] at $x = 0.055$ at $T = 0$, while $x = 0.15$ corresponds approximately to the optimum doping, where one achieves the highest-T$_c$ in LSCO. In order to explore the "stripe phase"[4] as extensively as possible, the optical conductivity $\sigma(\omega)$ of LSCO was determined between 295 and 30 K in seven single crystals with $x = 0$, 0.03, 0.05, 0.07, 0.10, 0.12, and 0.15 grown in different laboratories. The reflectivity $R(\omega)$ of the $a$-$b$ planes, which include the Cu-O planes where superconductivity takes place, was measured from either 20 or 40 cm$^{-1}$ to 20,000 cm$^{-1}$. Details of $R(\omega)$ in the far infrared are shown in the insets of Fig. 1. $\sigma(\omega)$, as extracted from $R(\omega)$ by Kramers-Kronig transformations,[9] is shown in Fig. 1 in the whole measured energy range.

The La$_2$CuO$_4$ sample in Fig. 1A is an antiferromagnetic insulator with a Néel temperature $T_N = 324$ K. Its spectrum shows three phonon peaks and additional weak peaks due to residual oxygen doping.[10] The absorption edge at high energy, here and in the other spectra of Fig. 1, is due to charge transfer (CT) between Cu and O ions.[11] The $\sigma(\omega)$ of LSCO with $x = 0.03$, not shown in the Figure, exhibits phonon lines as in Fig. 1A on the top of a broad background, which extends in the far- and mid-infrared. At $x = 0.05$ (Fig. 1B), just below the insulator-to-superconductor transition, a strong peak springs up at 250 cm$^{-1}$ as the temperature decreases. Its intensity overcomes by orders of magnitude the intensity of ordinary phonon lines, as those one may still glimpse at 350 and 680 cm$^{-1}$. The peak has the same $T$-dependence as a possible overtone at 500 cm$^{-1}$ and a broad absorption with a sharp edge at 580 cm$^{-1}$. The last broad band is similar to the $d$ band observed in Nd$_{2-x}$Ce$_x$CuO$_4$ and assigned to the excitation of charges self-trapped in the polar Cu-O lattice (polarons).[12] The intensity of all these features increases for $T\rightarrow 0$ through a transfer of spectral weight from higher-to-lower energy, which can hardly be appreciated on the logarithmic scale of the figure. A mid-infrared background (MIR band) also appears in Fig. 1B around 3000 cm$^{-1}$, in agreement with early experiments at room temperature.[11] Colored arrows, reported conventionally on the left of Fig. 1B, indicate the dc conductivity $\sigma_{dc}$ of the same sample, as obtained from the resistivity measured by the Van der Pauw technique[13] at the temperature of the corresponding spectrum. Here, as in other panels of the figure, the $\sigma_{dc}$ values are consistent with reasonable extrapolations of $\sigma(\omega)$ for $\omega \rightarrow 0$. The arrows also show that the free-carrier contribution peaked at $\omega = 0$ (Drude term) is much smaller, in the underdoped region, than the peak at 250 cm$^{-1}$ so that in no way the Drude term can be confused with the strong peak at $\omega \neq 0$. A similar situation has already been reported for a Bi$_2$Sr$_2$CuO$_6$ film with $T_c = 20$ K.[14] In the following, we shall focus on the intriguing far-infrared peak, of which we follow the evolution throughout the "stripe phase" of La$_{2-x}$Sr$_x$CuO$_4$.

The spectra of the superconductors with $x > 0.05$ in Fig. 1, including the one at optimum doping in Fig. 1E, are surprisingly similar to those of the sample in Fig. 1B with $x = 0.05$. A huge far-infrared peak at $\omega \neq 0$ is still observed, which becomes even stronger and softer. As a consequence, in Figs. 1C, 1D, 1E the first strong resonance is lower in energy than even the lowest transverse optical (TO) infrared-active phonon of La$_2$CuO$_4$, a mode with[15] $\omega_1^{TO} = 132$ cm$^{-1}$ at $x = 0$.

The systematic measurements reported in Fig. 1 show that unconventional absorption features are found in La$_{2-x}$Sr$_x$CuO$_4$ for a wide range of doping and temperature. Far-infrared peaks similar to those reported here were observed previously in a few superconducting samples of La$_{2-x}$Sr$_x$CuO$_{4+y}$, either oxygen-doped[16] or Sr-doped.[17] In the former case those peaks were tentatively assigned to polarons, in the latter case to disorder effects related to the Sr impurities. However, either for isolated[18] or for interacting polarons[19,20] the main absorption is predicted at frequencies higher than $\omega^{LO}$, where LO is a longitudinal optical mode of the polar lattice. As $\omega^{LO} > \omega_1^{TO}$, the peaks observed in Figs. 1C-1E at $\omega < \omega_1^{TO}$ cannot be attributed to conventional polaronic charges. On the other hand, the alternative interpretation in terms of spatial disorder may account for the increase with doping of the peak intensity, it may hardly explain its impressive increase for $T \to 0$.

In the following, we will show by several arguments that the strong far-infrared peaks of Fig. 1 are due to collective excitations of charge stripes.[21]

First of all, the spectra in Figs. 1C-1E are impressively similar in shape, energy, and temperature dependence to the ones reported[22] in systems like the hole-doped "ladder" $Sr_{14-x}Ca_xCu_{24}O_{41}$. This compound exhibits a spin gap and its infrared spectrum is interpreted in terms of collective modes of one-dimensional arrays of holes.[22] Also in the present layered cuprate, the extremely low frequencies $\omega_p$ of the peaks and their strength (see Fig. 1) point towards the excitation of massive, charged arrays coupled to a polar lattice. Let us assume that $\omega_p$ corresponds to the excitation of some vibrational mode of the whole array and that the vibrating mass is $M$. For the lowest-energy peak in Fig. 1 ($x = 0.12$, $T = 30$ K) $\omega_p \approx 30$ cm$^{-1} \approx \omega_{ph}/10$, where $\omega_{ph}$ is an average optical-phonon frequency. As $\omega_p$ should scale as $M^{1/2}$, one may infer that $M \approx 10^2\mu$, where $\mu$ is the reduced mass of a cell of $La_{2-x}Sr_xCuO_4$ vibrating at $\omega_{ph}$. This estimate of the stripe dimension in the $a$-$b$ plane ($\approx 10^2$ cells) is consistent with the available neutron scattering results, which give for the pinned stripes of $La_{1.28}Nd_{0.6}Sr_{0.12}CuO_4$ an average length of $\approx 20$ nm.[23] Therein, at higher temperatures the magnetic scattering peaks are lost, but inelastic neutron scattering still reveals stripe excitations with a threshold at about 3 meV ( 24 cm$^{-1}$).[23] Such situation should be similar to that of $La_{1.88}Sr_{0.12}CuO_4$ at any $T$, and that threshold value is in excellent agreement with the onset of the far-infrared absorption in Fig. 1D. It is also worth noticing that the lowest-$T$ peak energy is minimum for $x = 0.12$, indicating that stripes are most massive for this Sr concentration. Indeed, that doping value is close to $x = 1/8$, where the ordering length should increase due to commensuration effects.

A second argument comes out from the width of the far-infrared peak, which measures the lifetime of the excitation and/or an inhomogeneous spatial distribution of its characteristic frequency. Figure 2A shows the model proposed in Ref. 24 for commensurate charge stripes (in red) coexisting in $La_{1.28}Nd_{0.6}Sr_{0.12}CuO_4$ with antiferromagnetic (AF) stripes (in blue). One may reasonably assume that a close relation between AF and charge ordering holds also in the case of $La_{2-x}Sr_xCuO_4$ and for incommensurate stripes, even if on a shorter length and/or time scale. We report in Fig. 2B the halfwidth $HWHM_{IR}$ of the strongest observed far-infrared peak for each sample with $x > 0$. The star is an experimental point extracted from Ref. 16, under the assumption that the carrier density in $La_2CuO_{4.06}$ is roughly equivalent to that in $La_{2-x}Sr_xCuO_4$ with $x = 0.12$. Figure 2C is reproduced, instead, from Ref. 5. It shows for comparison the halfwidth $HWHM_{NS}$ of the peaks observed in the neutron scattering spectra of $La_{2-x}Sr_xCuO_4$ at 1.5 K and attributed to the AF stripes.[5] For the reader's convenience, the same guide to the eye is proposed in both panels of the Figure. A close resemblance in the doping dependence of the infrared and neutron scattering peaks is evident. Either the maximum width related to increasing fluctuations around the insulator-to-metal transition and the minimum around the commensurate value $x = 0.125$ are observed by both techniques. Figure 2 can hardly be explained in terms of alternative models[16,17] for the far-infrared peaks of LSCO, and provides strong support to the interpretation of the present data in terms of charge stripe excitations.

Finally, one should check that the far-infrared peaks are observed in a range of doping and temperature which coincides approximately with the "stripe phase" of the $T$-$x$ diagram of La$_{2-x}$Sr$_x$CuO$_4$ (see the inset of Fig. 3). We have then subtracted[12] the phonon contribution $\sigma_{ph}(\omega)$ and the background due to the bands $d$, MIR, and CT from $\sigma(\omega)$, as reconstructed from a fitting procedure, thus obtaining

$$\sigma^*(\omega) = \sigma(\omega) - \sigma_{ph}(\omega) - \sigma_d(\omega) - \sigma_{MIR}(\omega) - \sigma_{CT}(\omega). \qquad (1)$$

For $x = 0.15$ we have also subtracted the tail of the narrow Drude contribution, as reconstructed by a fit. The resulting, far-infrared, $\sigma^*(\omega)$ is thus suitable to calculate the spectral weight of the anomalous peaks, independently of their shape. As usual, this area is calculated in terms of an effective number of carriers per cell

$$n_{eff} = \frac{2m^* V}{\pi e^2} \int_{\omega_1}^{\omega_2} \sigma^*(\omega) d\omega. \qquad (2)$$

Therein, $m^*$ is conventionally assumed to be the free electron mass, $V$ is the volume of the cell, and $\omega_1$ is the lowest measured frequency. The cut-off frequency $\omega_2 > \omega_p$ is fixed, for any $x$, where $\sigma^*(\omega)$ has decreased to negligible values. The lower limit $\omega_1$ is also such as to exclude most of the narrow Drude term in the samples with $x < 0.15$. The resulting $n_{eff}$ is reported for all doped samples in Fig. 3 as a function of temperature. The error bars are largest for the spectra where $R(\omega) \approx 1$ and take into account the uncertainties involved in the procedure related to Eq. (1). As one may expect, at all temperatures $n_{eff}$ increases with $x$. Simple linear extrapolations to $n_{eff} = 0$ (dashed lines) provide the temperatures $T^*$ reported as red squares in the inset of Fig. 3. $T^*$ then represents here, for any $x$, a qualitative prediction of the temperature where the charge structures responsible for the far-infrared peaks should disappear. As shown in Fig. 3, those $T^*$ values are consistent with the cross-over region (shaded area) between the "stripe phase" and the normal metallic phase, as obtained from all the pseudogap temperatures collected by different techniques and reported in Ref. 7. Therefore, in addition to their energy and width, also the intensity of the far-infrared peaks consistently indicates that we have observed charge stripes in La$_{2-x}$Sr$_x$CuO$_4$ on the fast time scale of infrared spectroscopy.

1. J. Zaanen, O. Gunnarsson, *Phys. Rev. B* **40**, 7391 (1989).

2. V. J. Emery, S. A. Kivelson, *Physica C* **209**, 597 (1993).

3. S. Andergassen, S. Caprara, C. Di Castro, M. Grilli, *Phys. Rev. Lett.* **87**, 56401 (2001), and references therein.

4. J. M. Tranquada, B. J. Sternlieb, J. D. Axe, Y. Nakamura, S. Uchida, *Nature* **375**, 561 (1995).

5. M. Fujita *et al.*, *Phys. Rev. B* **65**, 64505 (2002).

6. A. Bianconi *et al.*, *Phys. Rev. Lett.* **76**, 3412 (1996).

7. For a review, see T. Timusk, B. Statt, *Rep. Prog. Phys.* **62**, 61 (1999).

8. H. Takagi *et al.*, *Phys. Rev. B* **40**, 2254 (1989).

9. A. Lucarelli *et al.*, *Phys. Rev. B* **65**, 054511 (2002).

10. G. A. Thomas, D. H. Rapkine, S. W. Cheong, L. F. Schneemeyer, *Phys. Rev. B* **47**, 11369 (1993).

11. S. Tajima *et al.*, *J. Opt. Soc. Am. B* **6**, 475 (1989).

12. S. Lupi *et al.*, *Phys. Rev. Lett.* **83**, 4852 (1999).

13. L. J. Van der Pauw, *Philips Research Rep.* **13**, 1 (1958).

14. S. Lupi, P. Calvani, M. Capizzi, P. Roy, *Phys. Rev. B* **62**, 12418 (2000).

15. R. T. Collins, Z. Schlesinger, G. V. Chandrashekar, M. W. Shafer, *Phys. Rev. B* **39**, 2251 (1988).

16. R. P. S. M. Lobo, F. Gervais, S. B. Oseroff, *Europhys. Lett.,* **37**, 341 (1997).

17. T. Startseva, T. Timusk, M. Okuya, T. Kimura, K. Kishio, *Physica C* **321**, 135 (1999).

18. D. Emin, *Phys. Rev. B* **48**, 13691 (1993).

19. J. Tempere, J. T. Devreese, *Phys. Rev. B* **64**, 104504 (2001).

20. S. Fratini, P. Quemerais, *J. Phys. IV France* **9**, Pr10, 259 (1999).

21. The broader contribution at higher energies (*d* band) can be explained in terms of the excitation of a single charge of the array. In the presence of a moderate or strong coupling with the lattice, this process corresponds to the photoionization of a polaronic charge.[12] At *x* = 0.5, where the *d* band is better resolved, good fits were obtained at all temperatures by using a large-polaron model for the optical conductivity.[18,19]

22. T. Osafune, N. Motoyama, H. Eisaki, S. Uchida, S. Tajima, *Phys. Rev. Lett.* **82**, 1313 (1999).

23. J. M. Tranquada, N. Ichikawa, S. Uchida, *Phys. Rev. B* **59**, 14712 (1999).

24. X. J. Zhou *et al.*, *Science* **286**, 268 (1999).

25. We thank M. Nuñez-Regueiro for giving two of us (M. O. and P. C.) hospitality in his laboratory for the resistivity measurements.

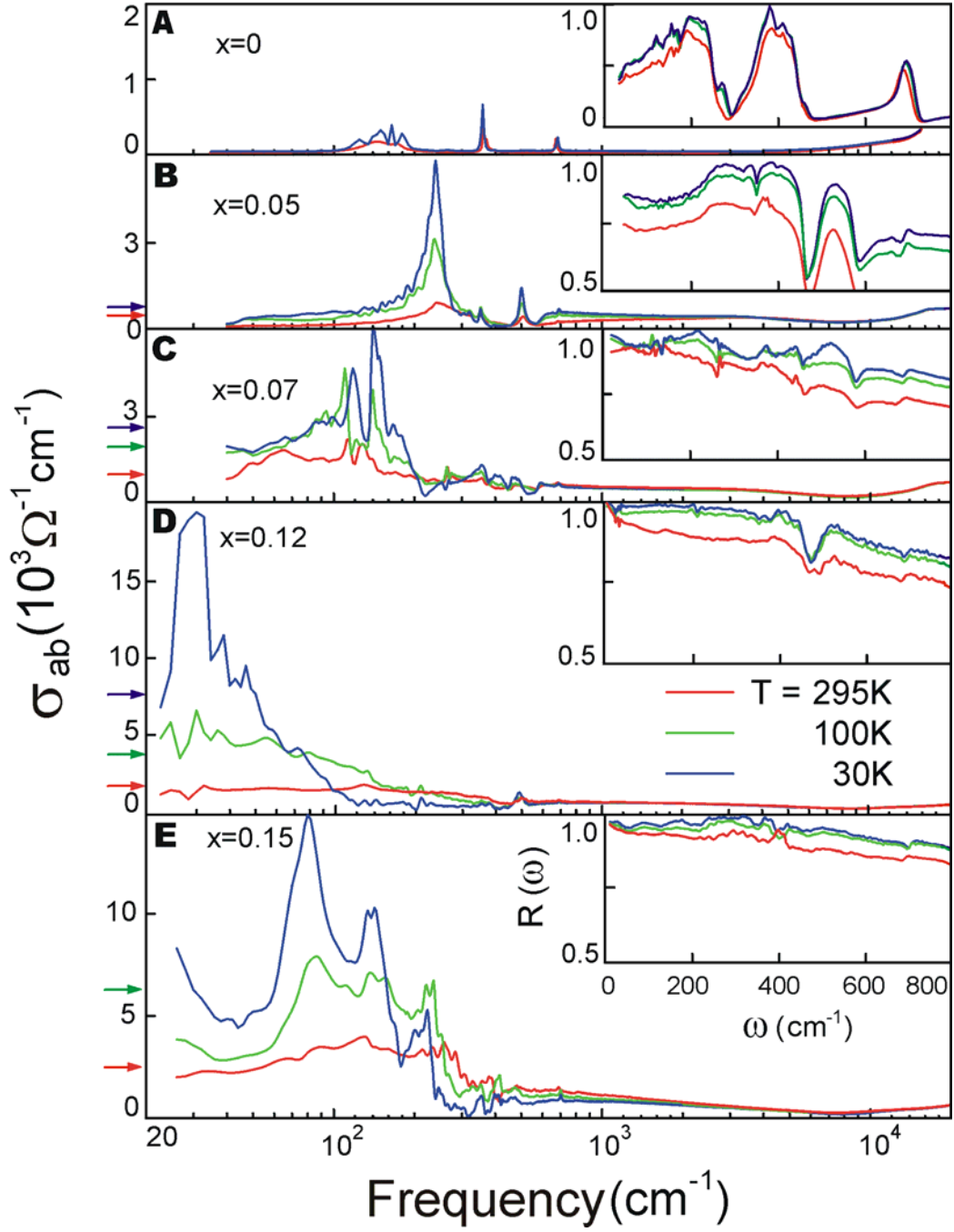

Figure 1. Optical conductivity of the *a-b* (Cu-O) planes for five La$_{2-x}$Sr$_x$CuO$_4$ single crystals with hole doping increasing from top to bottom, at different temperatures. Raw reflectivity data are shown in the insets for the far infrared range. Colored arrows mark the dc conductivity values at the temperature of the corresponding spectrum.

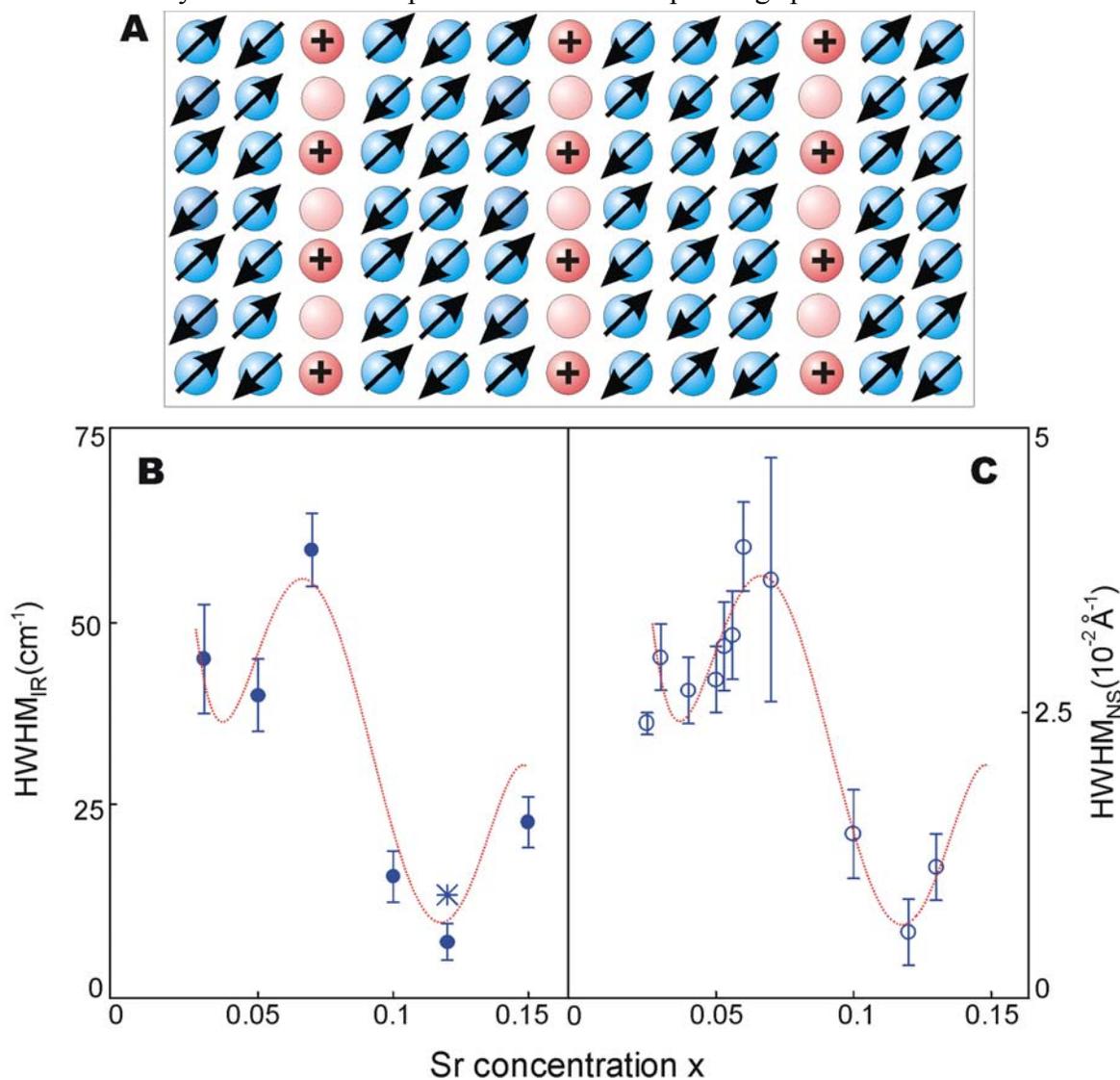

Figure 2. Comparison between the present infrared data and the neutron scattering results of ref. 5, based on the coexistence of charge stripes and antiferromagnetic domains. **A**, The model of Ref. 24 for the Cu-O planes at *x* = 1/8. The O ions are not shown, while the Cu

spins are marked by arrows and the excess holes by crosses. **B**, Halfwidth at half maximum of the strongest far-infrared peak in the 30 K spectrum, as a function of doping. The star is a point at 45 K from Ref. 16. **C**, Halfwidth at 1.5 K of the neutron-scattering peak due to the AF stripes, from Ref. 5.

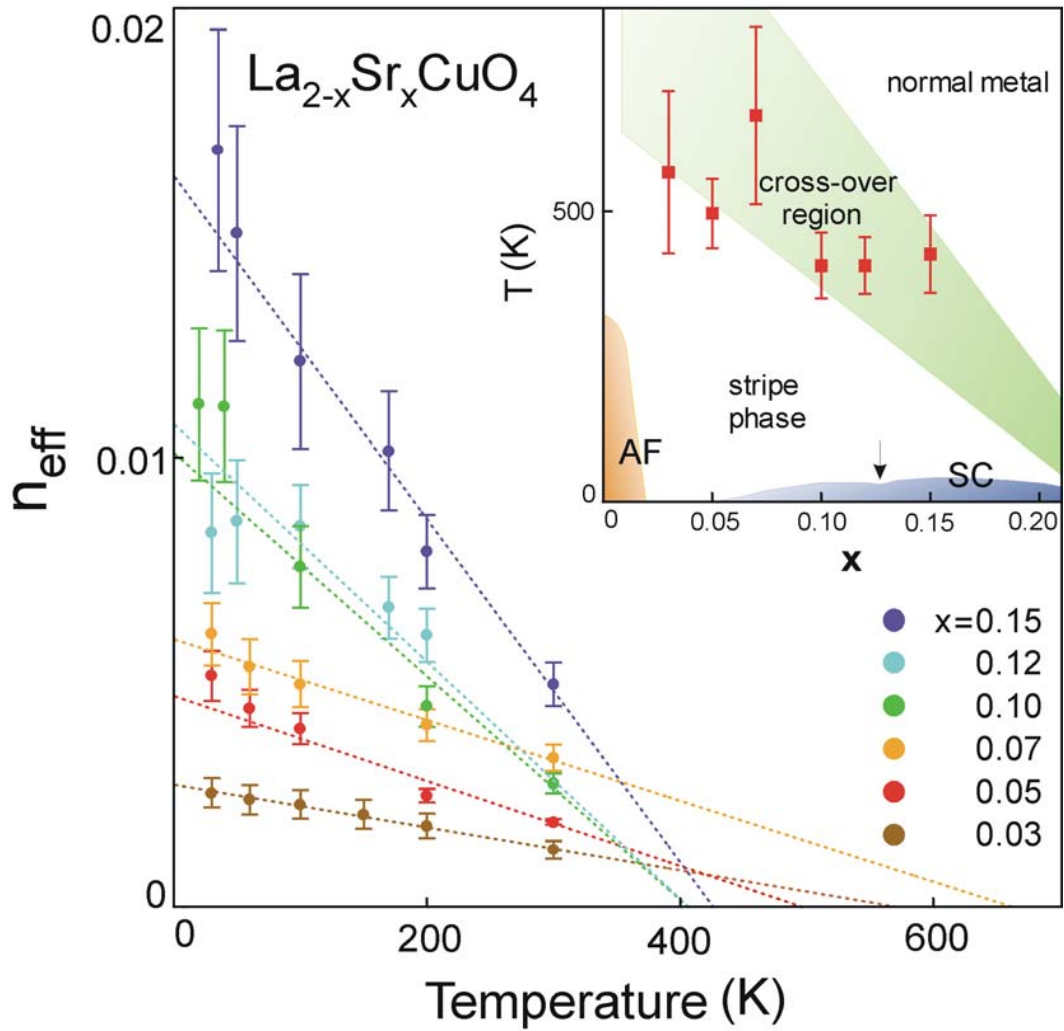

Figure 3. Effective number of carriers $n_{eff}$ (see Eq. 2) in the far-infrared peaks of Fig. 1, for different $x$, as a function of temperature. The linear least-square fits (dashed lines), once extrapolated to $n_{eff} = 0$ provide the temperatures reported in the inset by red squares. The green area marks the cross-over region between the "stripe phase" and the normal metallic phase, from the $T^*$ data reported in Ref. 7. AF is the antiferromagnetic, SC the superconducting phase of $La_{2-x}Sr_xCuO_4$. The arrow marks the minimum in $T_c$ at the commensurate doping $x = 1/8$.